\pdfoutput=1
\documentclass[twocolumn]{aastex62}

\usepackage{color}
\usepackage{amssymb}
\usepackage{amsthm}
\usepackage{multirow}

\received{*** **, ****}
\revised{*** **, ****}
\accepted{*** **, ****}
\submitjournal{ApJ}

\shorttitle{Dust in Compaction}
\shortauthors{Chen et al.}
\begin{document}

\title{Dust Temperature of Compact Star-forming Galaxies at $ z \sim 1-3$ in 3D-{\it HST}/CANDELS}

\correspondingauthor{Guanwen Fang}
\email{wen@email.ustc.edu.cn, xkong@ustc.edu.cn}

\author{Zuyi Chen}
\affiliation{School of Mathematics and Physics, Anqing Normal University, Anqing 246011, People's Republic of China}
\affiliation{CAS Key Laboratory for Research in Galaxies and Cosmology, Department of Astronomy, University of Science and Technology of China, Hefei 230026, People's Republic of China}
\affiliation{School of Astronomy and Space Science, University of Science and Technology of China, Hefei 230026, People's Republic of China}

\author{Guanwen Fang}
\altaffiliation{Guanwen Fang and Zuyi Chen contributed equally to this work.}
\affiliation{School of Mathematics and Physics, Anqing Normal University, Anqing 246011, People's Republic of China}

\author{Zesen Lin}
\affiliation{CAS Key Laboratory for Research in Galaxies and Cosmology, Department of Astronomy, University of Science and Technology of China, Hefei 230026, People's Republic of China}
\affiliation{School of Astronomy and Space Science, University of Science and Technology of China, Hefei 230026, People's Republic of China}

\author{Hongxin Zhang}
\affiliation{CAS Key Laboratory for Research in Galaxies and Cosmology, Department of Astronomy, University of Science and Technology of China, Hefei 230026, People's Republic of China}
\affiliation{School of Astronomy and Space Science, University of Science and Technology of China, Hefei 230026, People's Republic of China}

\author{Guangwen Chen}
\affiliation{CAS Key Laboratory for Research in Galaxies and Cosmology, Department of Astronomy, University of Science and Technology of China, Hefei 230026, People's Republic of China}
\affiliation{School of Astronomy and Space Science, University of Science and Technology of China, Hefei 230026, People's Republic of China}

\author{Xu Kong}
\affiliation{CAS Key Laboratory for Research in Galaxies and Cosmology, Department of Astronomy, University of Science and Technology of China, Hefei 230026, People's Republic of China}
\affiliation{School of Astronomy and Space Science, University of Science and Technology of China, Hefei 230026, People's Republic of China}

\begin{abstract}

Recent simulation studies suggest that the compaction of star-forming galaxies (SFGs) at high redshift might be a critical process, during which the central bulge is being rapidly built, followed by quenching of the star formation. To explore dust properties of SFGs with compact morphology, we investigate the dependence of dust temperature, $T_{\rm{dust}}$, on their size and star formation activity, using a sample of massive SFGs with $\log (M_{\ast}/M_{\odot})  > 10$ at $1 < z < 3$, drawn from the 3D-{\it HST}/CANDELS database in combination with deep {\it Herschel} observations. $T_{\rm{dust}}$ is derived via fitting the mid-to-far-infrared photometry with a mid-infrared power law and a far-infrared modified blackbody. We find that both extended and compact SFGs generally follow a similar $T_{\rm{dust}}-z$ evolutionary track as that of the main-sequence galaxies. The compact SFGs seem to share similar dust temperature with extended SFGs. Despite the frequent occurrence of AGNs in compact SFGs, we do not observe any effect on dust caused by the presence of AGN in these galaxies during the compaction. Our results disfavor different ISM properties between compact and extended SFGs, suggesting that a rapid and violet compaction process might be not necessary for the formation of compact SFGs.

\end{abstract}


\section{Introduction} \label{sec:intro}

The presence of massive quiescent galaxies (QGs) and star-forming galaxies (SFGs) with extremely small sizes in both observations and simulations (so-called red and blue ``nuggets'', respectively;
\citealt{Dokkum2008,Damjanov2009,Barro2013,Dekel2014, Zolotov2015,Tacchella2016a,Tacchella2016}) has spurred extensive discussion about a possible ``compaction and quenching'' evolutionary scenario.
During the compaction process, galaxies would acquire significant amount of central stellar mass in short timescale, followed by the cessation of star formation.
Evidences of such a two-step quenching mode have been found in local and distant galaxies \citep{Fang2013, Barro2013, Fang2015,  Barro2017, Wang2018, Woo2018}.
Possible mechanisms accounting for the compaction include mergers or/and violent disk instabilities, inducing gas inflows and resulting in active star formation in the central region \citep{Hopkins2008, Dekel2014, Zolotov2015, Tacchella2016}.
When the compact center of a galaxy is being rapidly built, the galaxy enters a starburst phase associated with short gas depletion timescale ($230^{+90}_{-120}$ Myr; \citealt{Barro2016}) as well as the triggering of active galactic nuclei (AGNs)/stellar feedback that initialize the quenching process of the galaxy \citep{Cai2013,Zolotov2015}.
Therefore, in massive SFGs where gas replenishment is usually suppressed \citep{Tacchella2016}, the compaction will end up with the formation of quiescent galaxies \citep{Tacchella2016, Barro2017}.

Nevertheless, it remains to be explored which mechanisms drive the compaction.
Recent studies reveal a dust-obscured nature for the compact star-forming galaxies \citep[cSFGs; e.g.,][]{Barro2014}, which would be the first step of the proposed ``compaction and quenching" scenario.
These galaxies appear to be dusty according to their locations on the popular {\it UVJ} diagram with many of them being detected by {\it Herschel}.
Furthermore, a much larger AGN fraction is found in cSFGs than in extended star-forming galaxies (eSFGs) and QGs along the ``compaction and quenching'' sequence, suggesting that AGNs might play a role during the compaction and quenching process \citep{Barro2014,Kocevski2017}. But how AGNs help to shape the physical properties and fueling process of their hosting cSFGs still remain to be fully explored \citep{Kocevski2017}.

The infrared (IR) emission of active SFGs primarily traces the dust-obscured star formation.
Dust properties (e.g., the dust mass and temperature), as usually constrained by a modified blackbody fitting to the FIR spectral energy distribution (SED), are linked to the physical conditions within which star formation occurs.
Thus, dust in the interstellar medium (ISM) provides a unique perspective to study properties of host galaxies such as gas content and the star formation efficiency \citep{Magnelli2014,Schreiber2016,Schreiber2018,Ma2019,Liang2019a}.
Besides star formation, AGN activity may also contribute to dust heating.
A harder radiation field may lead to a more efficient dust heating and thus a higher dust temperature.
This is particularly relevant for cSFGs which have a frequent occurrence of AGNs \citep{Barro2014,Kocevski2017}.
Therefore, the AGN effect may be uncovered through the analysis of dust content.

By investigating dust properties of cSFGs and eSFGs as well as the potential effect of AGNs on those, the work presented here is an attempt to gain new insights into the physical mechanisms involved in the compaction process.
Based on the wealth of multi-wavelength photometry collected from the {\it HST}, {\it Spitzer}, and {\it Herschel} observations in the Cosmic Assembly Near-infrared Deep Extragalactic Legacy Survey fields (CANDELS; \citealt{Grogin2011, Koekemoer2011}), we determine the dust properties of a large sample of massive SFGs at $1 < z < 3$ and constrain the nature of cSFGs.
We present the galaxy sample and their multi-wavelength data in Section~\ref{sec:data}, while their $T_{\rm dust}$ and star formation rate (SFR) are derived in Section~\ref{sec:cal}.
Our results, including a comparative analysis of the dust properties between extended and compact galaxies, are presented in Section~\ref{sec:result}, followed by a summary in Section~\ref{summary}.

Throughout the paper, a \cite{Chabrier2003} initial mass function (IMF) and a standard $\Lambda$CDM cosmology with $\Omega_{\rm m} = 0.3,\  \Omega_{\rm{\Lambda}} = 0.7 $ and $H_{\rm{0}} = 70\  \rm{km~s}^{-1}~\rm{Mpc}^{-1}$ are adopted.

\section{Data and Sample} \label{sec:data}

Four of the CANDELS fields are selected to conduct this work, where both robust rest-frame optical size measurements and deep mid-IR (MIR) to far-IR (FIR) photometries are available, i.e., the Great Observatories Origins Survey (GOODS) Northern and Southern fields (GOODS-N and GOODS-S), the Cosmic Evolution Survey field (COSMOS), and the All-wavelength Extended Groth Strip International Survey fields (AEGIS).
These valuable sizes measurements and infrared observations ensure a census of dust temperature variation during the compaction process of SFGs at the peak of the cosmic star formation history.

\subsection{3D-{\it HST} Survey}

Our sample is primarily based on data products from the 3D-{\it HST} spectroscopic survey \citep{Brammer2012} in combination with the CANDELS program \citep{Grogin2011,Koekemoer2011}.

We use the astrometry and photometry from the 3D-{\it HST} {\em Release v4.1} \citep{Skelton2014} but adopt the updated physical parameters from the {\em Data Release v4.1.5} \footnote{\url{http://3dhst.research.yale.edu/Data.html}} \citep{Momcheva2016} to select massive galaxies with $\log (M_\ast/M_{\odot}) > 10$ at $1 < z < 3$.
Only those with {\tt use\_phot = 1} in the 3D-{\it HST} catalogs are considered, i.e., galaxies with reliable photometric measurements.
We adopt the ``best'' galaxy redshift, $z_{\rm{best}}$, created by \citet{Momcheva2016}.
This $z_{\rm{best}}$ is determined by combing the grism redshift, if robust, with the spectroscopic and photometric redshifts compiled by  \cite{Skelton2014}.
Using $z_{\rm{best}}$, \citet{Momcheva2016} re-derive other auxiliary stellar population parameters, rest-frame colors, and SFRs from spectral energy modeling with FAST \citep{Kriek2009}.

Furthermore, the circularized radius, $r_{\rm{e}}$, is derived using $r_{\rm{e}} = R_{\rm{e}} \times \sqrt{q}$, where $R_{\rm{e}}$ is  the effective radius along the major axis and $q$ the axis ratio, as measured by \cite{Wel2014} from CANDELS {\it HST}/WFC3 imaging mosaics with GALFIT \citep{Peng2002}.
We correct them to the rest-frame optical size at 5000 \AA{} using Equation (2) in \cite{Wel2014} to achieve a uniform measurement.
We exclude galaxies with inaccurate or no size measurement, i.e., with {\tt Flag = 2 or 3} in the catalog of \cite{Wel2014}.

In total, $6303$ galaxies satisfy our mass and redshift cuts as well as {\tt use\_phot = 1}, $5700$ of which have robust size measurement and would be further considered in the next section.

\subsection{MIR to FIR Photometry and X-ray Data}
\label{sub:IR_phot}

The four CANDELS fields used here are all covered by the PACS Evolutionary Probe (PEP\footnote{\url{http://www.mpe.mpg.de/ir/Research/PEP/DR1}}; \citealt{Lutz2011,Elbaz2011, Magnelli2013}) survey at 100, and 160 $\mu$m (and 70 $\mu$m in GOODS-S) and by the {\it Herschel} Multi-tiered Extragalactic Survey (HerMES\footnote{\url{http://hedam.lam.fr/HerMES}}; \citealt{Roseboom2010, Roseboom2012, Oliver2012}) at 250, 350, and 500 $\mu$m.
Our MIR to FIR photometry is retrieved from the PEP and HerMES catalogs, which were extracted at the positions of the {\it Spitzer}/MIPS 24 $\mu$m sources.

To identify AGNs in our sample, we use X-ray observations from the {\it Chandra X-ray Observatory} ({\it Chandra}; \citealt{Weisskopf2000}).
The X-ray source catalogs for 2 Ms exposure of the {\it Chandra} Deep Field-North \citep{Xue2016} and 7 Ms exposure of the {\it Chandra} Deep Field-South \citep{Luo2017} are adopted for the GOODS-N and GOODS-S fields, respectively.
In these two catalogs, AGNs are already identified via several X-ray-based criteria \citep{Xue2016,Luo2017}.
For the COSMOS and AEGIS fields, we use the X-ray source catalogs from the {\it Chandra} COSMOS Legacy survey \citep{Civano2016, Marchesi2016} and the AEGIS-X Deep survey \citep{Brightman2014, Nandra2015}, respectively.
AGNs in these two fields are identified by either an intrinsic X-ray luminosity of $L_{\rm{0.5-7 keV}} \ge 3 \times 10^{42}\ \rm{erg\ s^{-1}}$ or an intrinsic photon index of $\Gamma \le 1.0$ (obscured AGNs), which are consistent with the criteria adopted by \cite{Xue2016} and \cite{Luo2017} in the GOODS-N and GOODS-S fields, respectively.
Because $L_{\rm{0.5-7 keV}}$ is not given in the catalogs of the COSMOS and AEGIS fields, we converse the provided $L_{\rm{2-10 keV}}$ to $L_{\rm{0.5-7 keV}}$ via
 \begin{equation}\label{eq:Lx}
 L_{\rm{0.5-7 keV}} = L_{\rm{2-10 keV}}/0.72,
 \end{equation}
assuming an intrinsic photon index of $\Gamma = 1.8$.

For each field, we use the PEP catalog as the reference one, and subsequently cross-match it with the 3D-{\it HST} catalog, the HerMES catalog, and the X-ray source catalog utilizing the method described below.
Briefly, for every source in the reference catalogue (i.e., the PEP catalogue), we search for the nearest counterpart within an optimised matching radius.
Here, we consider a range of matching radii from 0\farcs1 to 5\arcsec\ with a step of 0\farcs1, among which the optimised one is set to the largest radius satisfying the following two criteria: (1) the fake matched fraction, defined as the ratio of the fake matched number (see below) to the successful matched number, is smaller than 5\%; (2) the differential successful matched number, i.e., the new successful matched number with increasing matching radius, is larger than the differential fake matched number.
To estimate the fake matched number given two catalogs, we shift one of them by 30\arcsec\ in random directions 16 times and take the average of the 16 ``successful'' matched numbers as the fake matched number.
Clearly, the optimised matching radius depends on the number densities of the matched catalogs, and thus varies between catalogs even for the same field.
However, we note that most of the resulting optimised matching radii are smaller than 1\farcs5.

Since both the 3D-{\it HST} catalogs and the X-ray source catalogs provide redshift information, we require that $z_{\rm{best}}$ from the 3D-{\it HST} catalogs and $z_{\rm{X}}$ (redshifts of optical counterparts of X-ray sources) from the X-ray source catalogs should be in good agreement, i.e., $\mid z_{\rm{best}} - z_{\rm{X}} \mid / z_{\rm{best}} \leq 0.2$ (only one source was excluded by this criterion).
In total, the cross-match yields a sample of 2403 galaxies with at least one {\it Herschel} measurement and a robust {\it HST}/size measurement.

Furthermore, to ensure an accurate measurement of dust temperature, we remove sources with less than three detections at rest-frame wavelength longer than 50 $\mu$m. After these selections, we obtain a MIR-to-FIR catalog including 521 galaxies, 109 of which have AGN counterparts.

\subsection{Sample Selection}
\label{sec:sample_sel}

\begin{figure*}
	\centering
	\includegraphics[width=\textwidth]{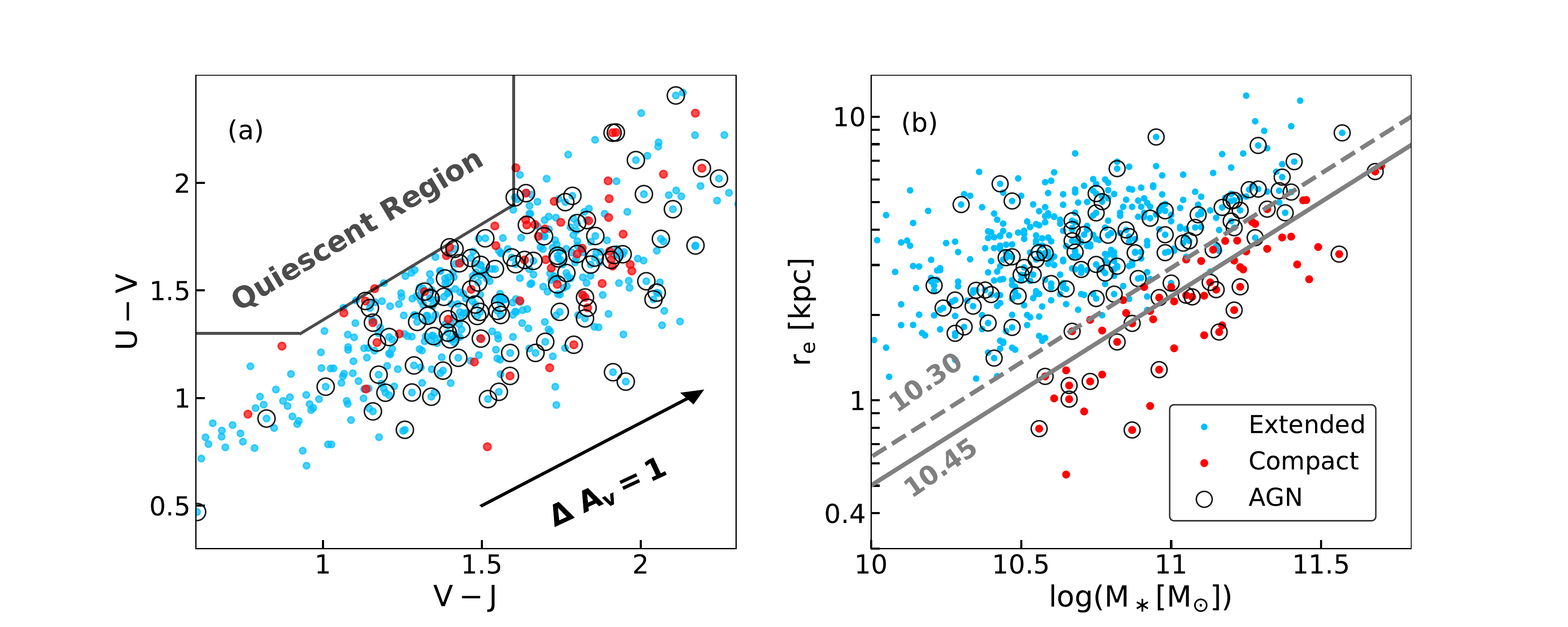}
	\caption{The {\it UVJ} diagram (left panel) and mass-size (right panel) distributions of our FIR-detected massive SFGs at $1 < z < 3$ with $\log(M_\ast/M_\odot) > 10$ and with an accurate measurement of their dust temperature.
	The \cite{Williams2009} criteria (black broken line) adopted to select SFGs are shown on the {\it UVJ} diagram, together with the direction and amplitude of extinction (vector) implied by the \cite{Calzetti2000} extinction law.
	The SFGs are further divided into extended SFGs (blue points) and compact SFGs (red points), according to the compaction criteria of $\log(M_\ast/r_{{\rm{e}}}^{1.5}/[M_\odot \ \rm{kpc^{-1.5}}]) = 10.45 $ (gray solid line) and 10.3 (gray dashed line) for galaxies at $1 < z \leq 2$ and $2 < z < 3$, respectively.
	Black circles indicate X-ray-identified AGNs. \label{fig:select}}
\end{figure*}

We use the {\it UVJ} criteria of \citet{Williams2009} to select SFGs amongst our catalog of 521 galaxies, resulting in 488 FIR detected massive SFGs with $\log (M_\ast/M_{\odot}) > 10$ at $1 < z < 3$.
To ensure a robust analysis of dust temperature, sources with catastrophic MIR-to-FIR SED fitting (see Section \ref{sec:cal_dt}) are also removed from our final sample.
We end up with a sample of \textbf{459} SFGs within which \textbf{95} galaxies are identified as X-ray AGN hosts.
Note that approximately $64\%$ of these galaxies have spectroscopic or grism redshift, which is key for accurate dust temperature measurements.

Finally, to identify cSFG from eSFGS, we use the compactness criteria defined by \cite{Barro2013}, i.e., $\Sigma_{\rm{1.5}} = \log(M_\ast r_{{\rm{e}}}^{-1.5}/[M_\odot \ \rm{kpc}^{\rm{{-1.5}}}])$.
Since the mass-size relation of SFGs evolves with redshift \citep{Wel2014}, cSFGs are identified as $\Sigma_{\rm{1.5}} \geq 10.3$ at $1 < z \leq 2$ \citep{Barro2013}  and $\Sigma_{\rm{1.5}} \geq 10.45$ at $2 < z < 3$ \citep{Barro2014}, respectively.
In total, there are 61 cSFGs.

Figure~\ref{fig:select} shows the distributions on the $UVJ$ and the mass-size diagrams for our final sample.
We observe a higher occurrence of X-ray AGNs in cSFG than in eSFGs, i.e., \textbf{37}\% versus 18\%, respectively.
This finding is consistent with the ubiquitous AGNs in cSFGs reported by \cite{Barro2016} and \cite{Kocevski2017}.
More than 80\% of our SFGs have $V - J$ colors redder than 1.2, a criterion used for selecting dusty SFGs in \cite{Spitler2014}.
Such red $V - J$ colors of our sample indicate that they are dust-obscured, which agrees with their {\it Herschel} detections.

\section{Dust Temperature and Star Formation rate}\label{sec:cal}

\subsection{Dust Temperature}\label{sec:cal_dt}

Following \cite{Casey2012} and using the corresponding IDL code {\tt cmcirsed}, we model the MIR to FIR (i.e., 24 to $500\ \mu \rm{m}$) SED of our SFGs, with a MIR power law plus a FIR greybody with a single dust temperature, $T_{\rm dust}$.
The power-law component accounts for the MIR emission of small clumps associated with hot dust or for AGN radiation, while the greybody accounts for the cold dust component emission \citep{Casey2012,Casey2014}.
For galaxies lacking sufficient FIR detection, the peak of the greybody emission will be poorly constrained, resulting in an unreliable estimate on $T_{\rm{dust}}$.
To reduce such effect, only galaxies with more than three data points at $\lambda_{\rm rest}>50\ \mu$m (i.e., where the greybody emission dominates) are included in our analysis (see Section \ref{sub:IR_phot}).
This extra selection criteria effectively translates into having for most of our galaxies more than four bands covering their MIR to FIR emission (76$\%$ have $\ge 5$ bands), providing us with a very accurate description of their infrared emission.

We fix the dust emissivity to a standard reference value of 1.5 \citep{Chapman2005,Casey2009}, leave the power-law slope ($0.5<\alpha<5.5$) as a free parameter \citep{Casey2012}.
We first perform an automatic fitting process assuming an optically thin dust model.
Then we visually inspect the best-fit SEDs for individual sources in order to ensure acceptable best-fit results.
For a few cases that the model fails to fit the data, we rerun the fit by manually adjusting (and fixing) $\alpha$ values.
For 25 galaxies, we are able to better constrain $T_{\rm dust}$ with manual fitting, and the new best-fit results are adopted.
However, there are still 29 of the original 488 FIR detected massive SFGs that fail to achieve an acceptable fit even though manual fitting process is implemented, due to either insufficient wavelength coverage that leads the peak emission unconstrained, or the fact that the data apparently deviate a lot from the assumption of MIR power law plus FIR greybody.
These galaxies with failed determination of $T_{\rm dust}$ are already excluded from our final sample (see Section \ref{sec:sample_sel}).

While dust temperature can be defined in different ways, such as mass-weighted or light-weighted, the light-weighted dust temperature derived as here using the FIR peak emission is what can be best constrained observationally \citep{Casey2014, Liang2019a}.
These light-weighted dust temperatures reflect the condition of the ISM surrounding massive and thus luminous star-forming regions.
They are also less dependent on the assumed underlying templates and allow for a direct comparison with previous works.

\subsection{Star Formation Rate}
\begin{figure*}
	\centering
	\includegraphics[width=\textwidth]{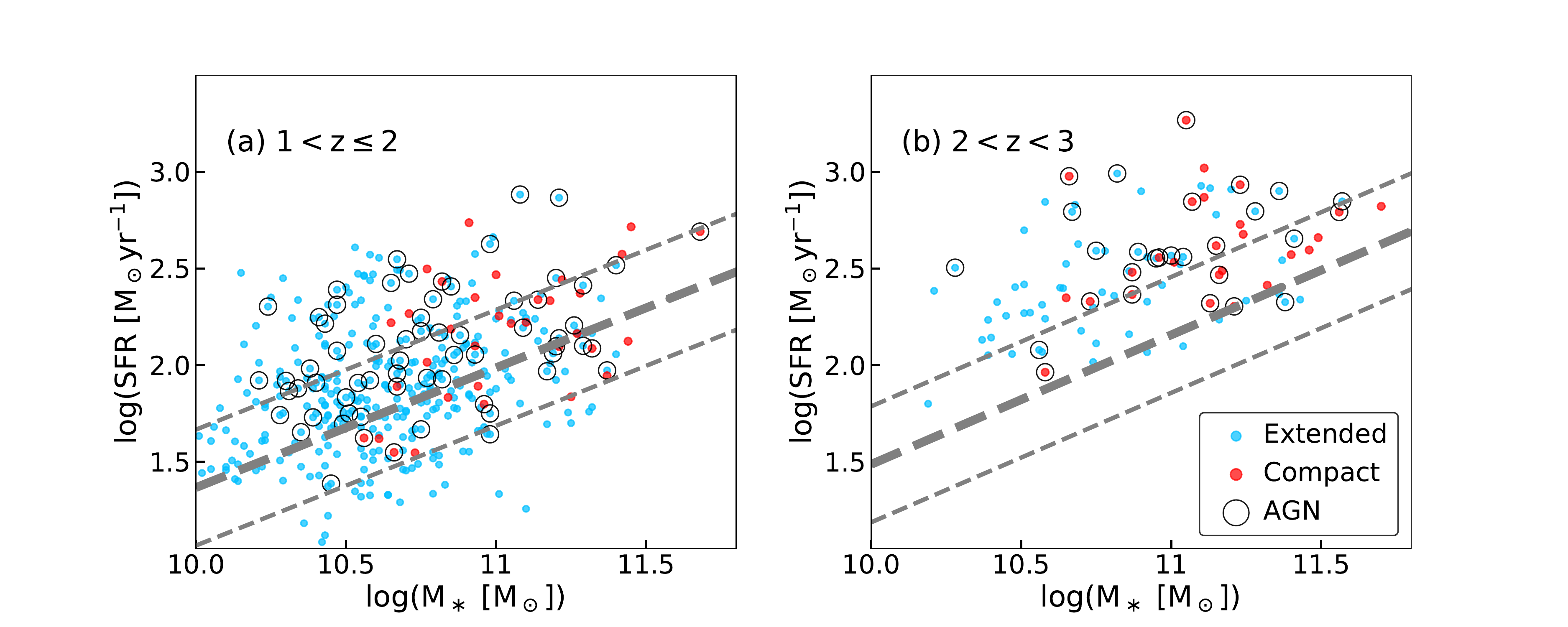}
	\caption{
	SFR--$M_\ast$ or main-sequence relations for our extended SFGs (blue points) and compact SFGs (red points) at $1 < z \leq 2$ (left panel) and $2 < z < 3$ (right panel), compared to the main-sequence relations of SFGs from \citet[][thick gray dashed lines plus thin ones for typical 0.3 dex scatter]{Whitaker2014}. The SFGs associated with X-ray-identified AGNs are marked with black circles. Compared with main-sequence galaxies, a significant number of our {\it Herschel}-detected compact and extended SFGs show elevated star formation activity, especially in the higher redshift bin.
    }
	\label{fig:ms}
\end{figure*}

The SFRs in the 3D-{\it HST} catalogs are derived using rest-frame UV luminosity, $L_{\rm{UV}}$, in combination with the {\it Spitzer}/MIPS 24 $\mu$m based total IR luminosities, $L_{\rm{IR}}$, as described in \cite{Whitaker2014}:
\begin{eqnarray}
\rm{SFR} & = & \rm{SFR}_{\rm{UV}} + \rm{SFR}_{\rm{IR}}, \nonumber\\
& = & 1.09\times 10^{-10} (2.2L_{\rm{UV}} + L_{\rm{IR}}).
\end{eqnarray}
We only keep the UV-based SFRs but recalculate $\rm{SFR}_{\rm{IR}}$ based on our MIR to FIR photometries.
We make use of the total IR luminosity, $L_{\rm{IR}}$, integrating the SED output by {\tt cmcirsed} from the rest-frame 8 to 1000 $\mu$m, and correct AGN contribution, $L_{\rm{IR,AGN}}$, with the method presented in \cite{Dai2018}.
Briefly, we convert the 2--10 keV luminosity of our X-ray AGNs to AGN emission at 6 $\mu$m using the well calibrated X-ray to MIR relation of \cite{Stern2015}
\begin{equation}
\log (L_{2-10\rm{keV}}/[\mathrm{erg\ s^{-1}}]) = 40.981 + 1.024x - 0.047x^2,
\end{equation}
where $x = \log (\nu L_{\nu}(6\ \mu \rm{m})/[10^{41}\ \rm{erg\ s^{-1}]})$. $L_{\rm{IR,AGN}}$ is then extrapolated by multiplying $\nu L_{\nu}(6\ \mu m)$ by a factor of 2.5, derived from the AGN template of \cite{Dai2012}.
We find generally small corrections for AGN contribution in our sample with a median value of $L_{\rm{IR,AGN}}/L_{\rm{IR}} \simeq 9\%$.
Note that for galaxies without X-ray AGN detection, the median of $L_{\rm{IR}}$ based on our FIR SED fitting is only larger by 0.07 dex than, and therefore broadly consistent with, that derived by \cite{Whitaker2014}.

The SFR--$M_*$ distributions for our SFGs are shown in Figure \ref{fig:ms} along with the main-sequence of SFGs from \cite{Whitaker2014}.
Compared to the main-sequence of SFGs, a significant number of our SFGs have enhanced star formation activity, especially in the high redshift bin.
By requiring FIR detection, our SFGs are biased to higher SFR.

\section{Results}\label{sec:result}

\subsection{Dust Temperature for Extended and Compact Galaxies}

\begin{figure}
	\centering
	\includegraphics[width=0.5\textwidth]{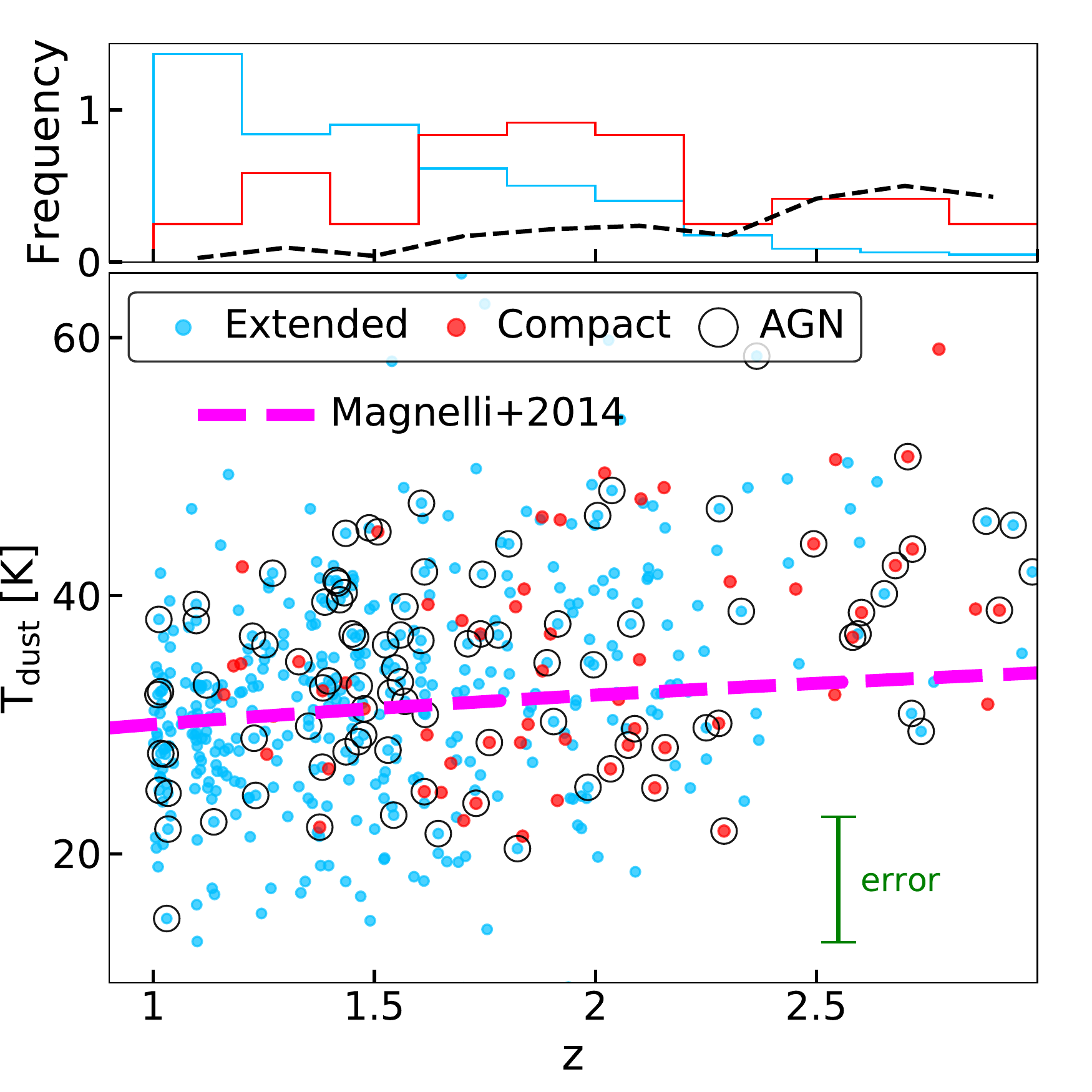}
	\caption{
        Top: normalized number distribution (i.e., the total area under the histogram is 1) of compact SFGs (red line) and extended ones (blue line), together with the fraction of compact SFGs at different redshift (black dashed line).
		Bottom: evolution of the dust temperature, $T_{\rm dust}$, with redshift for our extended SFGs (blue points) and compact SFGs (red points), compared to the overall redshift evolution of $T_{\rm{dust}}$ for the main-sequence galaxies (\citealt{Magnelli2014}; thick magenta dashed line).
		The SFGs associated with X-ray-detected AGNs are marked with black circles.
		The typical error of $\sim 5.5$ K for our sample is illustrated by the green error bar.
		Generally, the evolution of $T_{\rm{dust}}$ with redshift for both our extended and compact SFGs is similar to that of the main-sequence galaxies.
		\label{fig:tdust}}
\end{figure}

Dust temperature of SFGs is known to evolve with redshift in the sense that typical SFGs exhibit hotter dust temperature at high redshift than in the local universe, possibly due to a higher star formation efficiency or lower metallicity \citep{Magnelli2014, Schreiber2018, Liang2019a}.
Such an evolution cannot be neglected especially for our sample spanning a large redshift range.
Therefore, Figure \ref{fig:tdust} shows the dust temperature versus redshift for both eSFGs and cSFGs compared to that of main-sequence galaxies inferred by \cite{Magnelli2014}.
The median dust temperatures of eSFGs and cSFGs are also summarized in Table \ref{tab:mdust}.
We found that the number of cSFGs relatively to the number of eSFGs significantly decreases between $z\sim 2.5$ and $z\sim 1.5$ (upper panel of Figure \ref{fig:tdust}).
This is consistent with \cite{Barro2013} in which the decrease is explained by the fact that cSFGs are continuously formed at redshift 2--3 and gradually fade into QGs at later times.

\begin{table}
\caption{ The source number, dust temperature, and dust mass for eSFGs, cSFGs, and SFGs hosting AGNs in two redshift bins.}
\centering
\begin{tabular}{lccc}
\hline
\hline
Sample & $N$ & $T_{\rm dust}$/K & $\log(M_{\rm dust}/M_\odot)$ \\
\hline
$1 < z \le 2$ \\
eSFGs & 337 & 31.6 $\pm$ 0.5 & 8.1 $\pm$ 0.1 \\
cSFGs & 34 & 31.8 $\pm$ 1.8 & 8.4 $\pm$ 0.1 \\
AGNs & 67 & 33.3 $\pm$ 1.2 & 8.3 $\pm$ 0.1 \\
\hline
$2 < z < 3$ \\
eSFGs & 62 & 38.3 $\pm$ 1.9 & 8.5 $\pm$ 0.1 \\
cSFGs & 26 & 38.8 $\pm$ 2.9 & 8.7 $\pm$ 0.1 \\
AGNs & 28 & 38.7 $\pm$ 2.4 & 8.7 $\pm$ 0.1 \\
\hline
\end{tabular}
\tablecomments{For $T_{\rm dust}$ and $M_{\rm dust}$, the median is provided, together with uncertainties estimated via the bootstrapping method.}
\label{tab:mdust}
\end{table}

It is clear that both eSFGs and cSFGs generally follow the $T_{\rm{dust}} - z$ evolutionary track defined by the main-sequence SFGs presented in \cite{Magnelli2014}, especially for galaxies at $z\lesssim 2$.
From $z \sim 2.5$ to $\sim 1.5$, the dust temperature of cSFGs drops by $\mathbf{7.0 \pm 3.4}$ K. In the same redshift range, the dust temperature of eSFGs drops by $\mathbf{6.7 \pm 2.0}$ K.
The redshift evolution of dust temperatures for cSFGs and eSFGs are thus fully consistent within the uncertainties.

In the higher redshift bin, $T_{\rm{dust}}$ of both cSFGs and eSFGs are about 5~K higher than that inferred by \cite{Magnelli2014}.
This trend can, however, be explained by their higher SFRs as revealed in Figure \ref{fig:ms}.

Similarly, $T_{\rm{dust}}$ for AGN-hosting SFGs are also uniformly distributed around the $T_{\rm{dust}}$--$z$ track of main-sequence SFGs (see the open circle in Figure~\ref{fig:tdust} and Table~\ref{tab:mdust}).
No AGN effect on the temperature of the cold dust is observed for our sample.
However, further interpretation is limited by both the small sample size and the large uncertainty involved during the $T_{\rm dust}$ calculation.

Based on our FIR SED fits, we also infer dust masses ($M_{\rm{dust}}$) by extrapolating the rest-frame 850 $\mu$m emission of our galaxies and assuming a dust absorption coefficient of $\kappa_{850} = 0.15\ \rm{m^2\ kg^{-1}}$ \citep{Casey2012}. The resulting $M_{\rm{dust}}$ is given in Table \ref{tab:mdust}.
It can be seen that a huge amount of dust is enclosed in our SFGs, which is expected from their FIR detections.
Though cSFGs might have more dust compared to eSFGs, it should be treated with caution since the reliability of $M_{\rm{dust}}$ measurement is limited by the lack of data at longer wavelengths.
Our measurement is consistent with those of high-redshift dusty SFGs or sub-millimeter galaxies with $T_{\rm{dust}} \approx 20 - 50$ K and $\log(M_{\rm{dust}}/M_{\odot}) \approx 8 - 9.3$ \citep{Kovacs2006, Casey2012, Casey2014}.
Furthermore, the huge amount of dust in cSFGs also directly confirms the observations of \cite{Barro2014} that SFGs become more compact before they lose their gas and dust.

\begin{figure*}[htb]
    \centering
    \includegraphics[width=\textwidth]{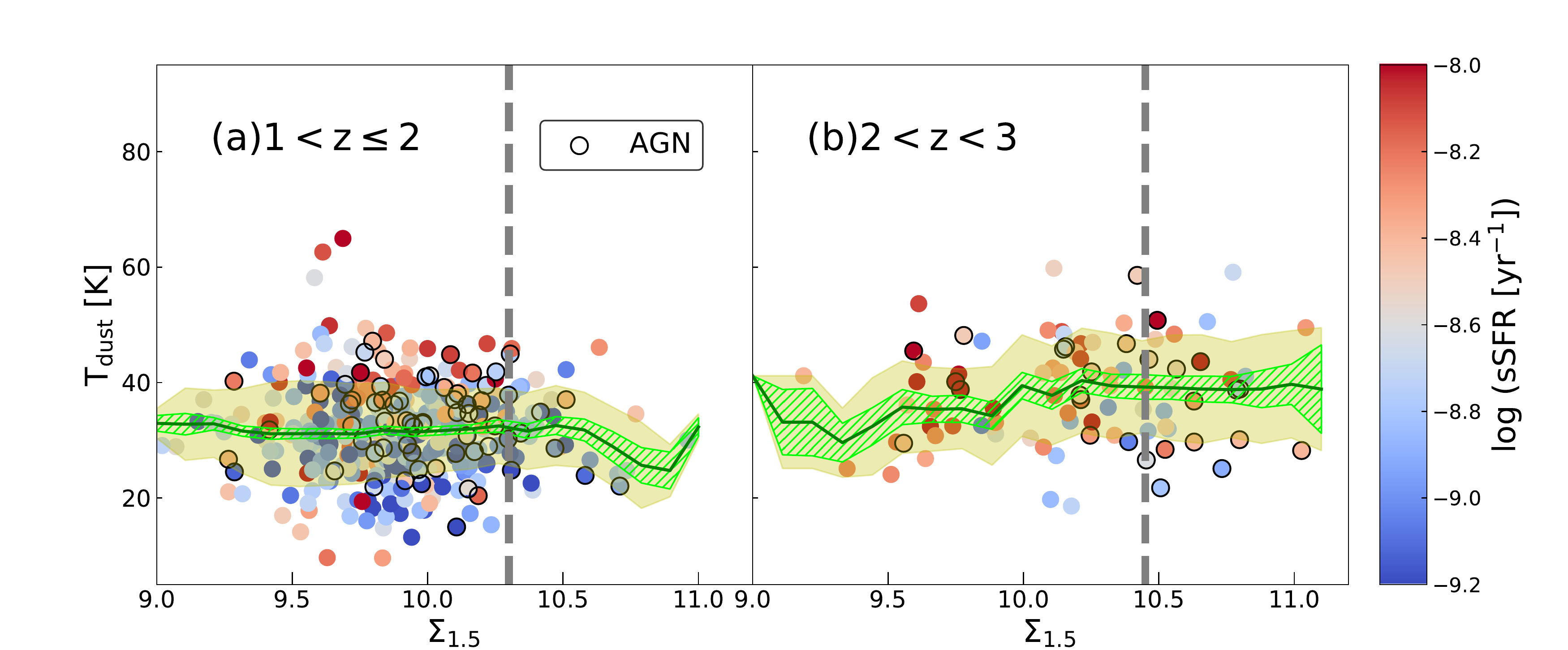}
    \caption{
    Relation between $T_{\rm{dust}}$ and $\Sigma_{\rm{1.5}}$ for our SFGs, color-coded by their sSFR, at $1 < z \leq 2$ (left panel) and $2 < z < 3$ (right panel). In each panel, the green solid line is the running medians of $T_{\rm{dust}}$ for each subsample, while the green hatched shadow indicates the $1\sigma$ uncertainties of the running medians estimated via the bootstrapping method. The yellow shadows show the running standard deviations of $T_{\rm{dust}}$ for the corresponding subsamples. The AGN hosts are marked by the black circles. The gray dashed lines indicate the cSFG criterion, i.e., $\Sigma_{1.5}=10.3$ and 10.5 for $1 < z \leq 2$ and $2 < z < 3$, respectively.}
  \label{fig:cmass-c}
\end{figure*}

To reveal the role of compactness in dust heating, we plot the $\Sigma_{1.5}$--$T_{\rm dust}$ relation, color-coded by specific star formation rates (${\rm sSFR}\equiv{\rm SFR}/M_{\ast}$), in Figure \ref{fig:cmass-c}.
Obviously, there is no significant trend between $\Sigma_{1.5}$ and $T_{\rm dust}$ in both redshift bins, while the Pearson linear (or Spearman rank) correlation coefficients are close to zero for both subsamples.
This result suggests a similar dust temperature for cSFGs and eSFGs, which is consistent with \cite{GomezGuijarro2019} in which the authors found that the ISM properties of cSFGs are similar to those of eSFGs located slightly above the main sequence.
Moreover, no significant feature of sSFR on the $\Sigma_{1.5}$--$T_{\rm dust}$ plane is observed in both redshift bins, although we will show in Section \ref{sec:sf} that cSFGs tends to have smaller sSFR compared to eSFGs at $2 <z<3$.

\subsection{Connection with Star Formation}
\label{sec:sf}

In this section, we focus on the effect of star formation activities on dust, with both the specific star formation rates  and the star formation rate surface densities ($\Sigma_{{\rm SFR}} = {{\rm SFR} / 2 \pi r_{\rm{e}}^2}$)\footnote{Note that $\Sigma_{\rm SFR}$ used in this work is defined by rest-frame optical size $r_{\rm e}$, the star formation-related size, however, is found to be smaller than the optical stellar size for SFGs at this redshift (e.g., \citealt{Elbaz2018,Fujimoto2018,Puglisi2019,Chen2020}).
Given the lack of the measurement of star formation-related sizes of these galaxies, we will use $\Sigma_{\rm SFR}$ defined by optical size to describe the concentration of star formation, but one should keep in mind that the derived values might be the lower limits of the true ones.}.

\begin{figure*}[htb]
	\centering
	\includegraphics[width=\textwidth]{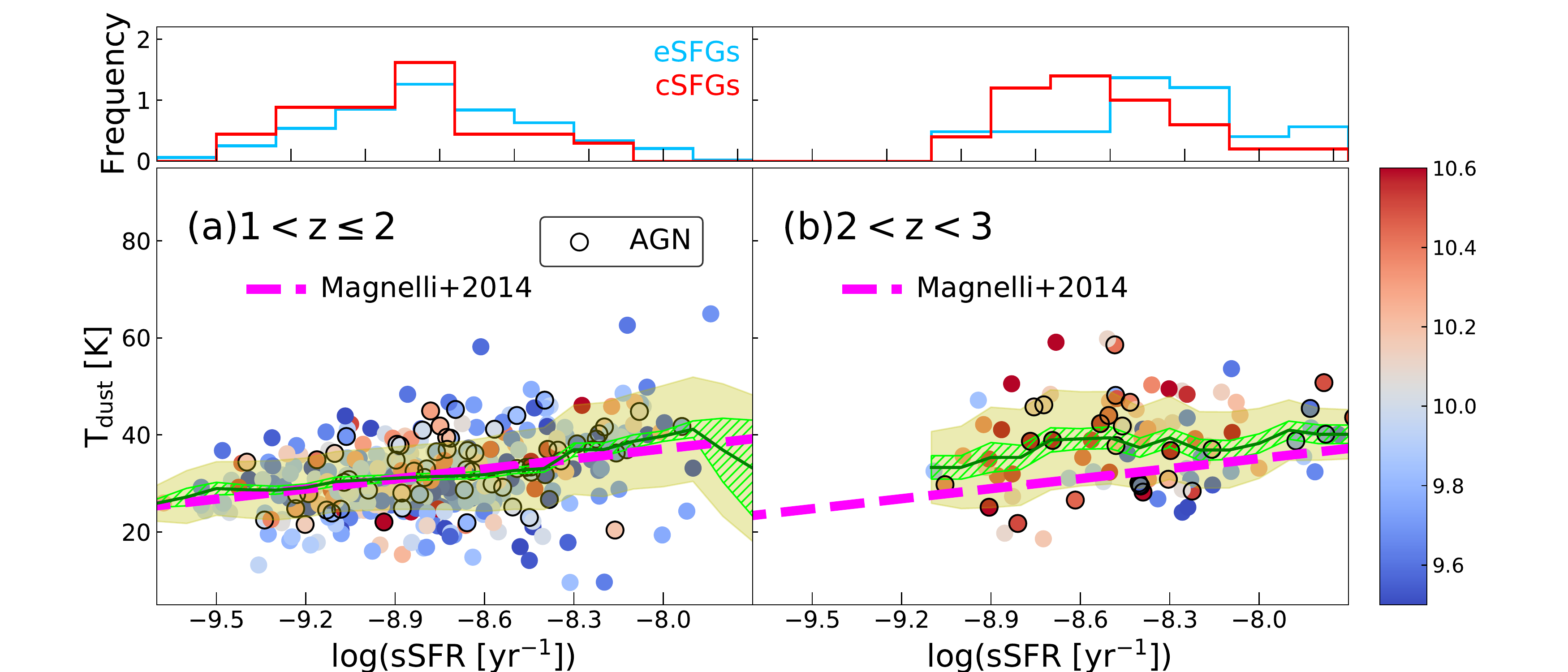}
	\includegraphics[width=\textwidth]{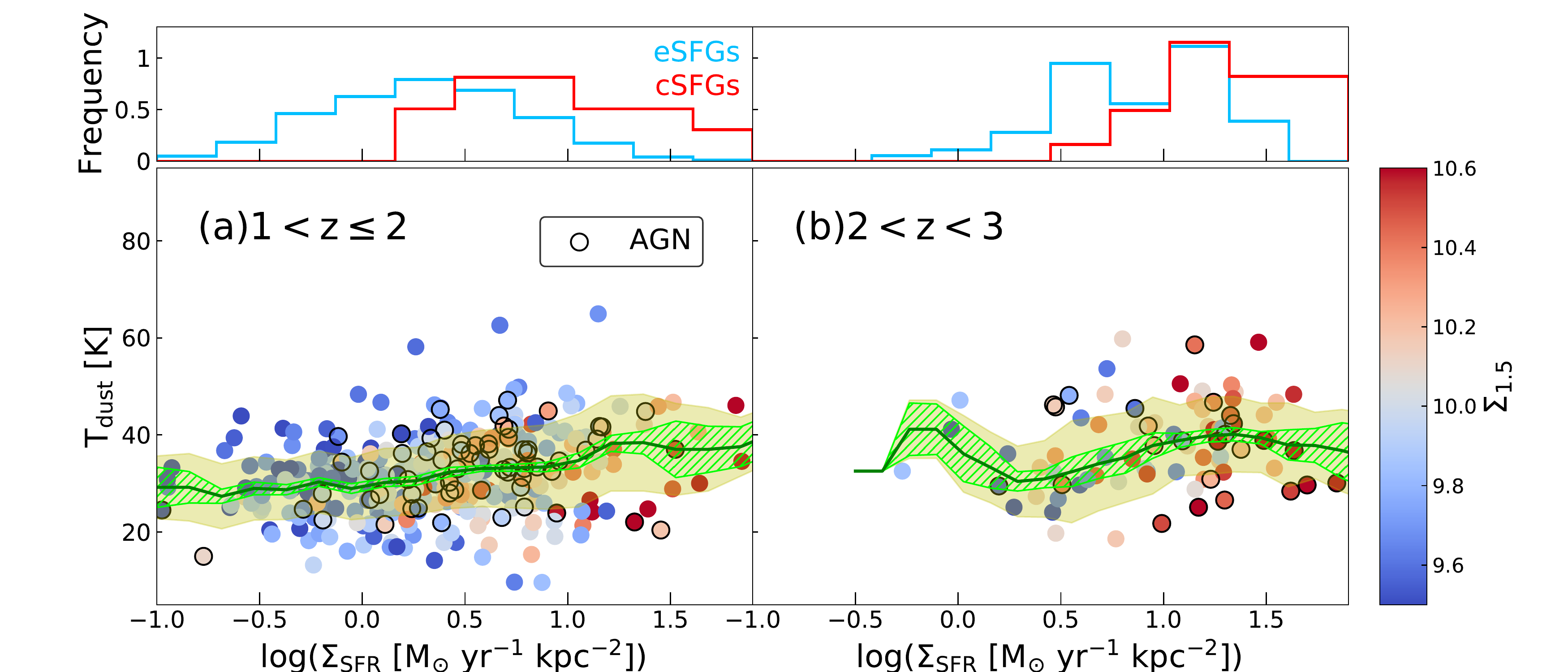}
	\caption{
	The $T_{\rm dust}$--sSFR (top) and $T_{\rm dust}$--$\Sigma_{\rm{SFR}}$ (bottom) relations for our SFGs, color-coded by their compactness ($\Sigma_{\rm{1.5}}$), at $1 < z \leq 2$ (left panel) and $2 < z < 3$ (right panel). The magenta dashed lines are the $T_{\rm{dust}}$--sSFRs relations at $z = 1.5$ (left panel) and $2.5$ (right panel) from \citet{Magnelli2014}. Other symbols are similar to those in Figure \ref{fig:cmass-c}. The normalized distributions of sSFR and $\Sigma_{\rm{SFR}}$ for eSFGs (blue) and cSFGs (red) are also given in the corresponding panel.}
    \label{fig:ssfr-c}
\end{figure*}

Dust temperature has been found to correlate with sSFR, better than other quantities like SFRs or $L_{\rm IR}$ \citep{Magnelli2014}.
In Figure \ref{fig:ssfr-c} we show the relation between $T_{\rm dust}$ and sSFR, color-coded by their compactness ($\Sigma_{\rm 1.5}$). The whole sample is divided into two redshift bins ($1< z \leq 2$ and $2 < z < 3$). The running medians of $T_{\rm dust}$ for two subsamples are shown in green solid lines, while the uncertainties of the running medians estimated via the bootstrapping method (green hatched shadows) and the running standard deviations (yellow shadows) are also plotted in Figure \ref{fig:ssfr-c}.

In each redshift bin, $T_{\rm dust}$ of our sample is roughly consistent with that of the main-sequence galaxies at similar redshift (magenta dashed line; \citealt{Magnelli2014}) within the large scatters.
At $1 < z \leq 2$, a weak positive correlation with a linear correlation coefficient of $\mathbf{r=0.34\pm0.06}$ between sSFR and $T_{\rm dust}$ is found, which supports that active star formation heats dust intensively. However, at $2 < z < 3$, due to the lack of small-$T_{\rm dust}$ SFGs in the low-sSFR regime, such a relation is more ambiguous, and $T_{\rm dust}$ seems to show no correlation with sSFR ($\mathbf{r=0.16\pm0.09}$).
The lack of small-$T_{\rm dust}$ SFGs with relatively low sSFR at $2 < z < 3$ might be due to the selection effect, i.e., our FIR selection bias to galaxies with more intense star formation compared to the main-sequence SFGs at similar redshift \citep{Whitaker2014}, as suggested by the right panel of Figures \ref{fig:ms} and the large void region below the \cite{Magnelli2014} relation at $z\sim2.5$ in Figure \ref{fig:tdust}.

In the low-redshift bin, SFGs with different compactness are well mixed, which is supported by the similar distributions of sSFR for eSFGs and cSFGs given in Figure \ref{fig:ssfr-c}.
However, both the distribution of individual galaxies and the comparison of the sSFR distributions between the eSFG and cSFG subsamples imply that more compact SFGs (redder color in Figure \ref{fig:ssfr-c}) seem to possess lower sSFR at $2 < z < 3$, which might be a signature of suppression of star formation. However, as shown in the right panel of Figure \ref{fig:ms}, our FIR selected galaxies form an SFR detection limit of $\sim 100~M_{\odot}$ yr$^{-1}$, which bias our sample to more star-bursting galaxies at low-$M_*$ end. Given that eSFGs in this subsample tend to be less massive than cSFGs,\footnote{Figure \ref{fig:ms} clearly shows that less cSFGs are observed at the low-$M_*$ ends in both redshift bins. In fact, our FIR selection should bias our sample to more compact galaxies due to their compact morphology and thus possibly higher IR surface brightness. Therefore, given an IR detection limit and a fixed SFR, cSFGs should be much easier to be detected, which conflicts with our observations. The physical reason for this conflict is still unknown and might be worth further study.} eSFGs thus might be more star-bursting (higher sSFR) than cSFGs. In other words, the observed lower sSFR of cSFGs at $2 < z < 3$ might result from the FIR selection combining with the lack of cSFGs at the low-$M_*$ end.
Further $M_*$-controlled comparison reveals that cSFGs and eSFGs have similar sSFR within the same $M_*$ bins, supporting that the apparently lower sSFR of cSFGs might be just a selection effect.

High-redshift cSFGs also seem to have slightly hotter dust than that predicted by the main-sequence SFGs \citep{Magnelli2014}.
This might suggest differences in the ISM conditions of these compact objects with respect to typical SFGs with similar sSFR.
However, considering the low number statistic and the aforementioned selection effect, this result should be treated with caution.

Similar plots for the $T_{\rm dust}$--$\Sigma_{\rm SFR}$ relation are also shown in Figure \ref{fig:ssfr-c}.
SFGs with greater compactness (gray to red points) have significantly larger $\Sigma_{\rm SFR}$ than eSFGs, suggesting that size shrinkage is more dramatic than SFR changes.
The linear correlation coefficients between $T_{\rm dust}$ and $\Sigma_{\rm SFR}$ are $\mathbf{0.30\pm0.06}$ and $\mathbf{0.24\pm0.08}$ for the $1<z\leq2$ and $2<z<3$ subsamples, respectively, indicating a weak relation in both redshift bins.

From Figures \ref{fig:ssfr-c}, we find that AGN hosts are distributed in a similar way as those without AGN. The two populations share the same dependence of $T_{\rm{dust}}$ on star formation activities, in terms of sSFR and ${\Sigma_{\rm{SFR}}}$.
Star formation thus might be the main heating source of dust in these galaxies.

In short, at $1<z\leq z$, cSFGs seem to be indistinguishable from eSFGs except for their compact stellar morphology, while an apparently suppression of star formation is observed for cSFGs at $2<z<3$, however, this might be a selection effect.

\subsection{Role of cSFGs}
\label{sec:role_cSFGs}

Given the current data, we only observed an apparently suppression of star formation for our selected cSFGs compared to eSFGs at $2<z<3$.
The role of these cSFGs in the picture of galaxy evolution is still unclear.

\cite{Elbaz2018} compiled a sample of $z\sim2$ \textit{Herschel}-selected galaxies, which resemble the massive end ($M_\ast\gtrsim 10^{11}~M_{\odot}$) of our high-redshift subsample,\footnote{Note that seven out of eight \textit{Herschel}-selected galaxies in \cite{Elbaz2018} have redshift of $\gtrsim 2$.} and found that galaxies located within the scatter of the main sequence tend to have compact star formation and short depletion timescales.
The authors argued that these galaxies could be at the last step of star formation.
Although we do not have measurement of star formation-related size for our sample, the most massive galaxies at $2<z<3$ located within the scatter of the main sequence also tend to have compact stellar morphology (see Figure \ref{fig:ms}), possibly suggesting a similar properties to those galaxies hidden in the main sequence in the \cite{Elbaz2018} sample.

Furthermore, based on observations of galaxies at $0.5<z<3$, \cite{GomezGuijarro2019} reported that cSFGs and eSFGs exhibit similar star formation efficiencies, while the ISM properties (e.g., CO excitation and local physical conditions of the neutral gas) of cSFGs are also similar to those of eSFGs located slightly above the main sequence.
Besides the above ISM properties, our results (i.e., Table \ref{tab:mdust} and Figure \ref{fig:cmass-c}) provide further supplement related to dust that no significant differences in $T_{\rm dust}$ and $M_{\rm dust}$ are observed between cSFGs and eSFGs for both redshift subsamples.

Therefore, for cSFGs in both redshift bins, because of their similarity to eSFGs, it is possible that these galaxies form their compact stellar cores via slow secular evolution rather than a rapid compaction process, as suggested by \cite{GomezGuijarro2019}.

\section{Summary}\label{summary}

In this work, we investigate how dust temperature depends on the compactness and the star formation activity, using a sample of massive high-redshift SFGs with FIR observations from {\it Herschel} in four of 3D-{\it HST}/CANDELS fields.
Our findings are summarized as follows.

\begin{itemize}

\item Dust temperature of both eSFGs and cSFGs agree, within the uncertainties, with the $T_{\rm{dust}}-z$ evolutionary track of the main-sequence SFGs found by \cite{Magnelli2014}.

\item No significant difference in dust temperature is found between cSFGs and eSFGs.

\item While similar sSFR is found for eSFGs and cSFGs at $1 < z \le 2$, cSFGs tend to show smaller sSFR at $2 < z < 3$, which results from our FIR selection and the lack of low-$M_*$ cSFGs.
Compared to eSFGs, cSFGs are observed to have larger $\Sigma_{\rm SFR}$, which is consistent with their remarkable size shrinkage.

\item $T_{\rm{dust}}$ seems to slightly correlate in a positive way with sSFR at $1 < z \le 2$, but such trends disappears at $2 < z < 3$, which might cause by FIR selection effects.

\item No difference of dust temperature is observed whether AGN activities are ignited or not, though AGNs have higher occurrence in cSFGs that is consistent with the literature \citep{Barro2013, Fang2015, Kocevski2017}.

\item Our findings, together with other ISM study in the literature, suggest similar ISM properties between eSFGs and cSFGs, which implies that a rapid and violent compaction process might be not necessary for the formation of cSFGs.
\end{itemize}

The intense star formation in the central bulge and strong AGN feedback may lead to extreme properties of the ISM in the compact phase.
However, our results disfavor different ISM properties in cSFGs compared to eSFGs, questioning the existence of a rapid compaction process to form such compact galaxies.
Our sample size is limited by both precise size measurement and the needs for multi-wavelength {\it Herschel} detection.
Future work is still required to disclose the detail physics of the dynamic ISM in the compact phase, in order to complete the picture of ``compaction and quenching'' scenario.

\acknowledgments

We are grateful to the anonymous referee whose constructive and detailed comments helped us improve this paper.
We thank Zhenyi Cai for help polishing the manuscript.
This work is supported by the National Natural Science Foundation of China
(NSFC, Nos. 11673004, 1320101002, 11433005, and 11421303) and the National Basic Research
Program of China (973 Program) (2015CB857004).
Z. Lin gratefully acknowledges support from the China Scholarship Council (No. 201806340211).
G.W.F acknowledge support from Yunnan young and middle-aged academic and technical
leaders reserve talent program (No. 201905C160039) and Yunnan Applied Basic Research Projects (2019FB007).

This work is based on observations taken by the 3D-HST Treasury Program (GO 12177 and 12328) with the NASA/ESA HST, which is operated by the Association of Universities for Research in Astronomy, Inc., under NASA contract NAS5-26555.
This research has made use of data from HerMES project (\url{http://hermes.sussex.ac.uk/}). HerMES is a Herschel Key Programme utilising Guaranteed Time from the SPIRE instrument team, ESAC scientists and a mission scientist.
The HerMES data was accessed through the Herschel Database in Marseille (HeDaM - \url{http://hedam.lam.fr}) operated by CeSAM and hosted by the Laboratoire d'Astrophysique de Marseille.

\bibliography{ms.bib}

\end{document}